\documentclass[12pt,preprint]{aastex}

\shorttitle{Precession of the White Dwarf}
\shortauthors{Tovmassian et al.}

\begin{document}

\title{Evidence of Precession of the White Dwarf in Cataclysmic Variables}

\author{Gaghik H. Tovmassian\altaffilmark{1,2} and Sergey V. Zharikov\altaffilmark{1}}
\affil{Observatorio Astron\'{o}mico Nacional, Instituto de Astronoma, UNAM, M\'exico. }
\email{\{gag,zhar\}@astrosen.unam.mx}
\and

\author{Vitaly V. Neustroev}
\affil{Computational Astrophysics Laboratory, National University of Ireland, Galway, Newcastle Rd.,
Galway, Ireland.}
\email{Vitaly.Neustroev@nuigalway.ie}

\altaffiltext{1}{For correspondence use: PO Box 439027, San Diego, CA 92143-9027, USA.}

\altaffiltext{2}{Visiting Research Fellow, CASS, UCSD}

\begin{abstract}
Cataclysmic  Variables (CV)  are close  binary systems,  in  which the
primary, the more massive star, is a white dwarf.  CVs usually exhibit
a number of periodicities, most of which are now understood.  However,
recently,  a new  phenomenon  was  discovered that  does  not fit  the
standard picture.   Two objects have  been discovered to  show periods
that are much longer than orbital,  and have no relation to it, either
in light  curves or in  radial velocity (RV) variations  measured from
spectroscopy.  Here,  we show that the precession  of the fast rotating magnetically
accreting white  dwarf can  successfully explain these  phenomena. The
theory of compact objects  predicts certain relations between the spin
and precession periods,  and our finding provides a  good test for the
theory  and  establishes  a  qualitative  model to  be  explored  both
theoretically and observationally.  Detection of precession can become
a  powerful tool  in searching  for the  internal properties  of compact stars,
which would be otherwise inaccessible for us.
\end{abstract}


\section{Introduction}

Cataclysmic Variables (CVs) are close binary systems, with typical orbital
periods of a few hours. The more massive component in the binary, the primary star,  is
the white dwarf (WD) which accretes matter, usually through the disk
formed around it. This produces number of periodic or quasi
periodical signals in the light of the object observed both
spectroscopically and photometrically \citep{Warner}. Usually these
signals are close to the orbital period (within 2-3\%), as in the case
of superhumps, either positive or negative \citep{2001PASP..113..736P} or much
shorter as in the case of Quasi-Periodic Oscillations (QPOs) 
and flickering, which are not periodic, but may
introduce complicated patterns in the power spectra for the short time scales.
In some cases, the final accretion occurs along the magnetic lines if
the white dwarf is magnetized. In cases when the magnetic field of the primary exceeds
$\approx10$ MG, the magnetosphere reaches beyond the disk radius, completely destroys it and
locks the spin of the WD with the orbital period. Such objects are called Polars
and they usually do not exhibit confusing periods.  If the magnetic field of the 
WD is not so strong, but  B$> 0.5$ MG,  the inner parts of the
accretion disk become truncated, but the rest of the disk is not necessarily destroyed.
However, the magnetic poles on the surface of the WD become a source of high energy
beams due to the shock created by the accreting matter on them (similar to Polars).
The beams functioning as a lighthouse  allows detection of WD's strictly periodic spin period,
which is not locked with the orbital period in this case. The
spin period is normally shorter than the orbital period and often
found to be about 10\% of the latter. The beams from magnetic poles can
irradiate other components of the binary system. Usually that creates
a complex modulation of light observed from such objects,
sub-classified as Intermediate Polars (IP).  In spite of these
complications, the period analysis of most IPs was a relatively
easy task \citep{1986MNRAS.219..347W,1996MNRAS.280..937N}.

Recently though, two objects were
found to show periods that could not be explained within existing
schemes.  \object{FS Aur}, a seemingly ordinary dwarf nova system with an 85.7
min spectroscopic period was found to exhibit 3.425 hour photometric
variations in its light curve \citep{Tov1}.  Extended
monitoring of the system over more than 10 years confirmed that the
long photometric period has a variable amplitude and can be
contaminated by other variability, but it remains stable and coherent
\citep{Neustroev3}.
In order to explain the long photometric period in \object{FS Aur}, \cite{Tov1}
invoked the IP model with a fast rotating and precessing, magnetically accreting white dwarf.
Analyses of the fast ULTRACAM photometry of  \object{FS Aur}  suggests
the  presence of  the quickly spinning white dwarf in the system \citep{Neustroev2}.

\object{HS2331+3905} was discovered later and
immediately attracted attention for having two unrelated spectral
periods \citep{Araujo04}. Contrary to any conventional logic,
it exhibits two very different spectral periods, a short one at the center, and a
longer one in the wings of its emission lines. The radial velocity
variations of the emission lines of CVs, thanks to the Doppler effect,
are the principal means of period determination in binary
stars. Luckily, \object{HS2331+3905} is of high inclination and shows eclipses
in its light curve, thus allowing an unambiguous determination of the
orbital period of 81.08 min. The second period ($\sim3.5$ hours), appearing
in the far wings of the emission lines, has no explanation to
date.
Besides its two spectroscopic periods, \object{HS2331+3905}
exhibits a number of periodic photometric variations and is proved to
contain a magnetic WD. The 67.2 sec spin period of the WD
defines this object as an IP \citep{Araujo04}.

%

This study is based on observations of these two CV systems that have
unexplained long periods, one in its light curve  and the other in radial
velocities.  Here, we report the
detection of the second long (non-orbital) period in the RV measurements of emission
lines of \object{FS Aur} that converges these two objects into a common case
and suggests an universal picture explaining their behavior.

\section{Observations and reduction of data}

The observations of \object{FS Aur} were made
at the 2.1 m telescope of San Pedro Martir Observatory in Mexico in December  2004.
During two nights, using the Boller \& Chivens  spectrograph,  we gathered 81 spectra 
with total 12.4 hours of coverage of \object{FS Aur} with 420 sec individual exposure times.
We obtained  54 spectra with FWHM 4.1\AA\ resolution around the H$\beta$
line and  additional 27 spectra redward of H$\alpha$.
Including readout time between exposures, each spectrum covers about 0.1 orbital 
phase and approximately 0.03 phases of the long photometric period.
The log of observations is presented in Table \ref{obslog}.  Arc lamp
exposures were obtained every 2 hours during long runs to insure
proper wavelength calibration. For the flux calibration spectroscopic
standard stars Feige67 and Feige110 were observed.
The reduction was done using IRAF\footnote{IRAF is the Image Reduction and Analysis Facility, a general
purpose software system for the reduction and analysis of astronomical
data. IRAF is written and supported by the IRAF programming group at
the National Optical Astronomy Observatories (NOAO) in Tucson,
Arizona. NOAO is operated by the Association of Universities for
Research in Astronomy (AURA), Inc., under cooperative agreement with
the National Science Foundation}
routines which included bias subtraction, weighted
extraction of the spectra, as well as wavelength and flux calibration.

The same telescope with the same instrument and slightly better resolution (FWHM 2.5\AA) were
used during earlier observations of  \object{HS2331+3905} in September 
2004 in a large multi-longitude, international campaign.
The rest of our observations as well as  a thorough
analysis of larger set of data collected for this object will be
presented in an upcoming paper by \cite{Gansicke}. In Table
\ref{obslog} there is an entry corresponding to the observation of
\object{HS2331+3905} used here.
Actually we
only use a trailed spectrogram of \object{HS2331+3905} obtained by stacking
about 70 spectra from  8 hours continuous exposures into one
two-dimensional image. The exposure times were also of 420 sec,
similar to the data obtained for \object{FS Aur}.
The spectra of both objects were
normalized to the continuum
before being stacked into two-dimensional trailed spectrum.



\section{The results}

\subsection{Evidence of the long period in the FS Aur spectra}


Usually FS Aur in its quiescence varies between  aproximately $15.^m4-16.^m2$ with
an average magnitude about 15.7(V) and shows a complex variability in the light
curve which  is a combination   of  the long term period 3.425 h, the orbital period
85.7 min and QPO at short time scales  \citep{Neustroev2, Tov1}. The spectrum is typical
for a dwarf nova with single-peaked strong and broad H, HeI emission lines and weak
HeII $\lambda$4686, CIII/NIII blend.  The emission lines  velocity variations clearly
show the orbital period of the system reported by \cite{Thorst}.

\begin{figure}
\epsscale{.75}
\plotone{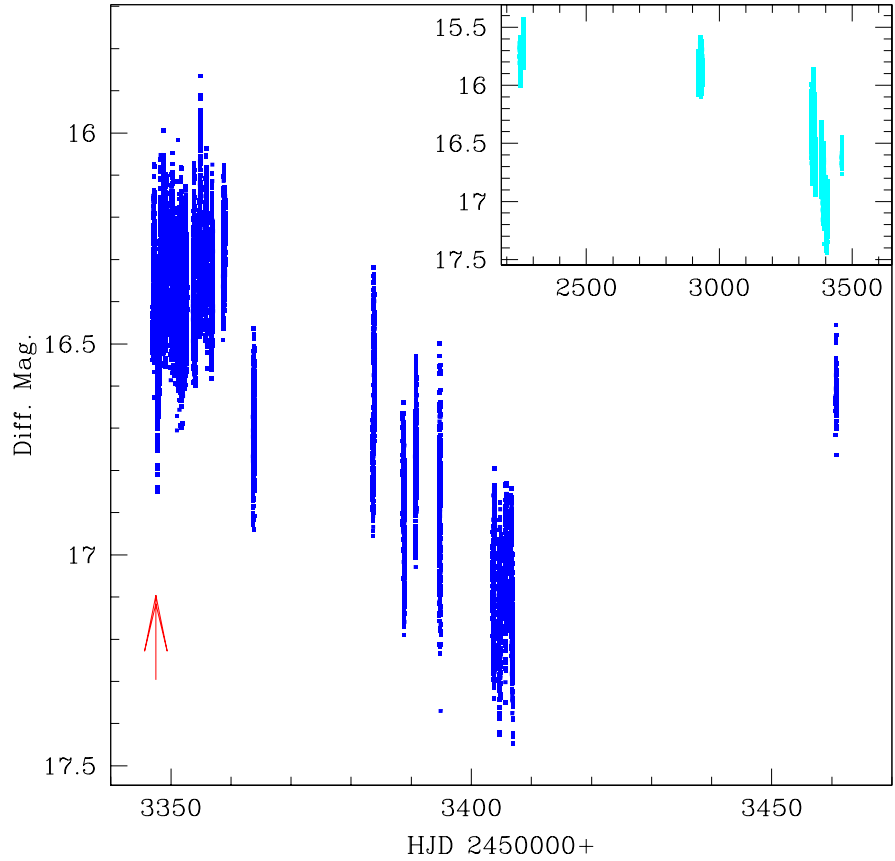}
\caption{The light curve of FS Aur. The spectroscopic observations reported in this paper were
performed during two nights marked with the arrow. The normal quiescence
brightness of FS Aur  is $\approx 15.7$(V)). The inset of the figure shows long term
light curve prior to our new observations where the object was in quiescence.
During the spectroscopy the object was already in a low state and shows an
anti-dwarf nova drop in brightness afterwards.\label{fig1}}
\end{figure}

During the new spectral observations presented here, \object{FS Aur} was found
to be more than half a magnitude fainter than expected for its quiescent
state.  It continued to decrease its brightness by about 1.5 magnitudes until the 
end of January 2005, after that coming back to the typical quiescent state.
In Fig.\ref{fig1} a long term
light curve is showing this trend. The AAVSO\footnote{www.aavso.org} data demonstrate
that the decrease in luminosity lasted about a few  months.
Such behavior is known in CVs as a VY Scl or anti-dwarf nova phenomenon.
VY Scl objects are a rare group of
CVs that undergo sudden and cyclical decrease of their luminosity, contrary to dwarf nova,
that are called so, for showing cyclical outbursts.
They are mostly concentrated in the 3--4 hour range of orbital periods, but recently shorter 
period systems have been found to exhibit the same kind of behavior \citep{dwcnc}.  
Also in recent years there were attempts to link VY Scl phenomenon with truncation of the inner 
disk by the magnetosphere of the WD \citep{Lasota,Tov2}. 
The detection of a VY Scl type drop in brightness alone is an indication that 
\object{FS Aur} is probably a magnetic CV.
The decreased brightness of the system, usually due to the drop in  mass accretion rate
and mainly affecting the continuum level, allowed us  to study faint wings of the lines with
clarity.

 At first glance, the new spectra of \object{FS Aur} looked similar to what was
seen in previous observations, namely single peaked emission lines with a
strong core whose sinusoidal RV variations define  the
orbital period. Generally it is  thought that the center of
the line originates at the outer edges of the accretion disk and is
often dominated by a bright, hot spot at the point of impact of
the stream of transfer matter with the disk. In order to measure
radial velocities close to the WD, which will better reflect the
orbital motion,  the double Gaussian method \citep{Schneider} is used.
The method allows for measurements of RVs in the far wings of
the lines, originating in the inner parts of the accretion disk, close
to the white dwarf. This technique involves convolving individual spectra with a double
Gaussian template. When the counts in the two Gaussian bandpasses are equal, then
the velocity of the emission line is assigned the value indicated by the midpoint between the two
Gaussians. By varying the width and separation of the Gaussians, the velocity variations of various
portions of the emission-line profiles is measured. This method is routinely used in study of
CVs in order to measure extreme line wings in conditions of low S/N and determine radial velocity
of the primary.

\begin{figure}
\epsscale{.75}
\plotone{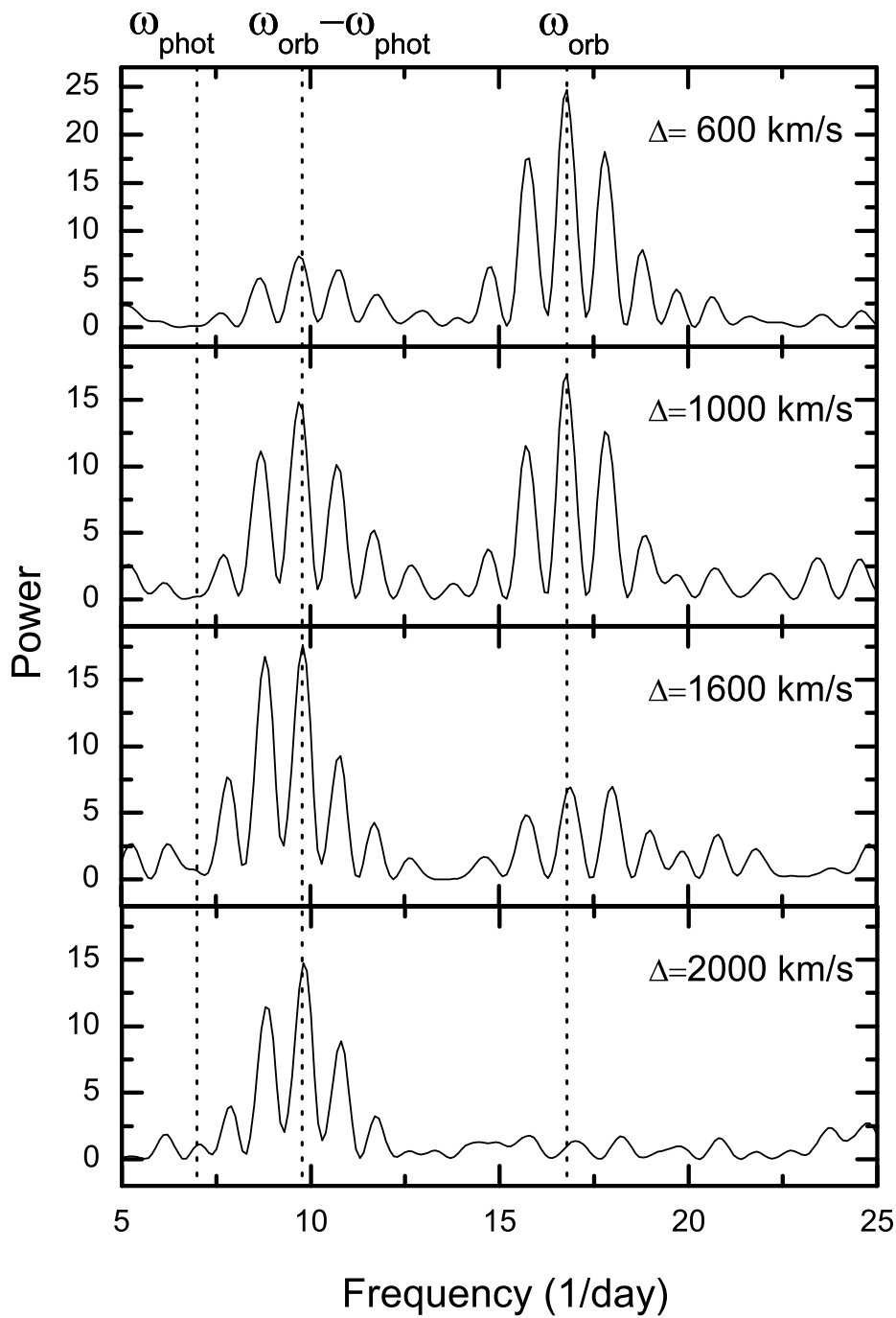}
\caption{The periodogram of radial velocity variations of the H$\beta$ line
of FS Aur, measured by the double Gaussian method. There were 81 spectra
measured, spanning 11.4 hours on two consecutive nights (~4 photometric
periods). In the upper panel, the period calculations (Scargle 1982) are
based on measurements with the 600 km/sec separation of the Gaussians, and
the power in the spectrum peaks at the orbital period. Increasing the
separation to 1000 km/sec produces the emergence of the second significant
period. At 1600 km/sec the second period becomes dominant and coincides
exactly with the beat period between orbital and photometric periods
observed in FS Aur.
\label{fig2}}
\end{figure}

\begin{figure}
\includegraphics[angle=0,scale=1.05]{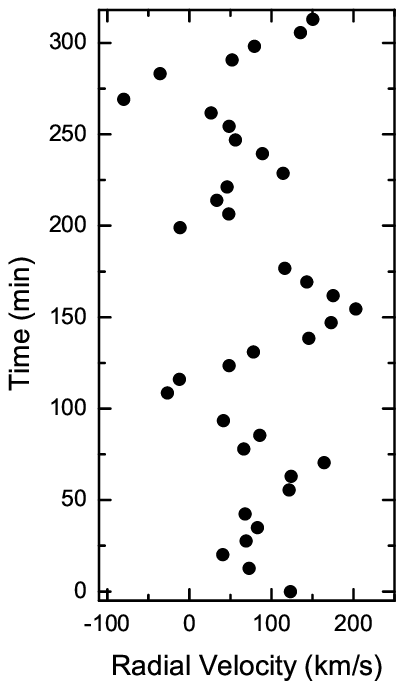}
\includegraphics[angle=0,scale=1.05]{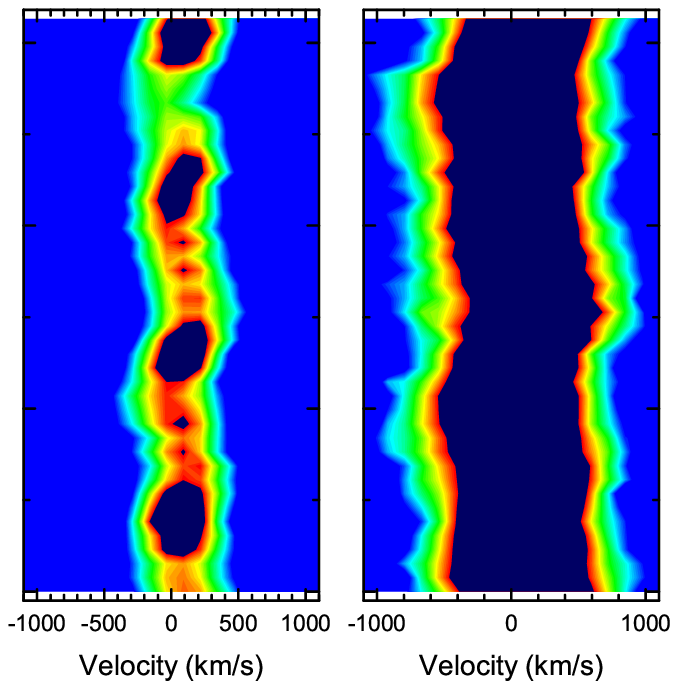}
\includegraphics[angle=0,scale=1.05]{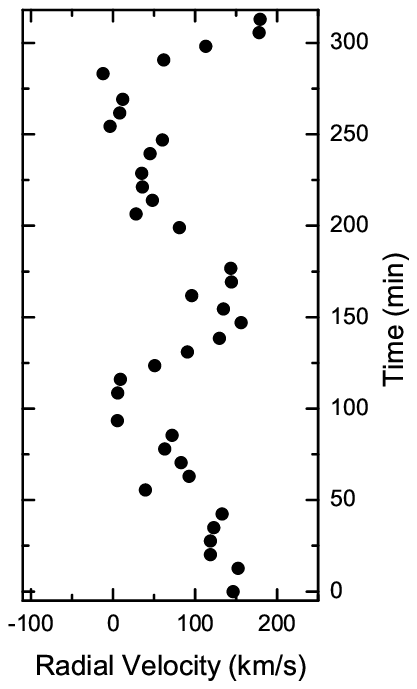}
\caption{The trailed spectrogram of H$\beta$ emission line of FS Aur made
of 37 individual spectra obtained during the first observing night,
are shown on two central panels with different contrast to emphasis different
components of the line. The trailed spectra are flanked by their corresponding RV curves.
The first panel on the left presents measurements of the center of the line with double Gaussian separation of 600 km/sec. The last panel presents measurements with 1600 km/sec separation. The emission line is modulated with two periods.
The central part  varies according to the orbital
period, while wings of the line better seen in the right panels are modulated
with the beat period (147 min) between the orbital and long photometric
period.
\label{fig3}}
\end{figure}

\begin{figure}
\epsscale{1.1}
\plottwo{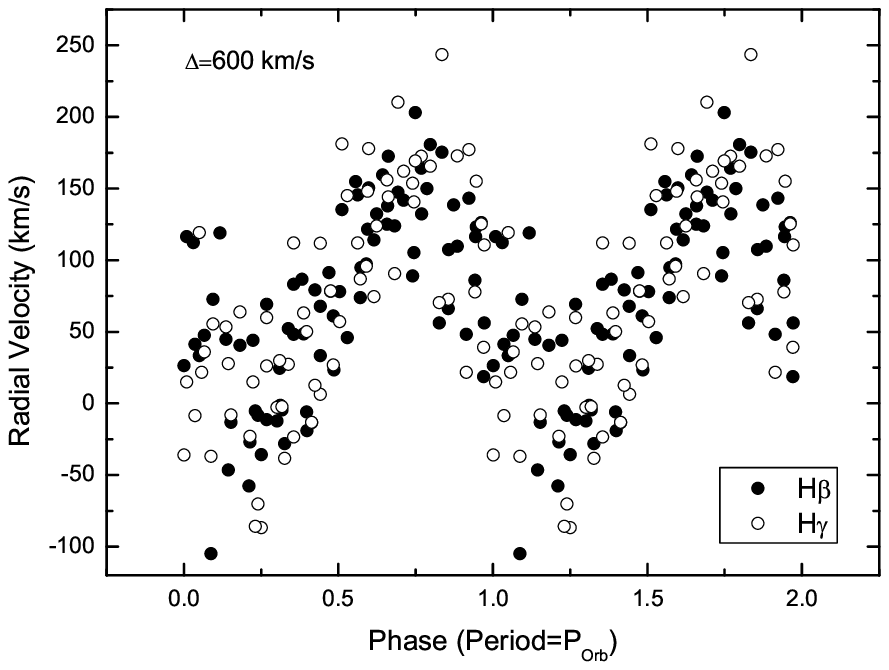}{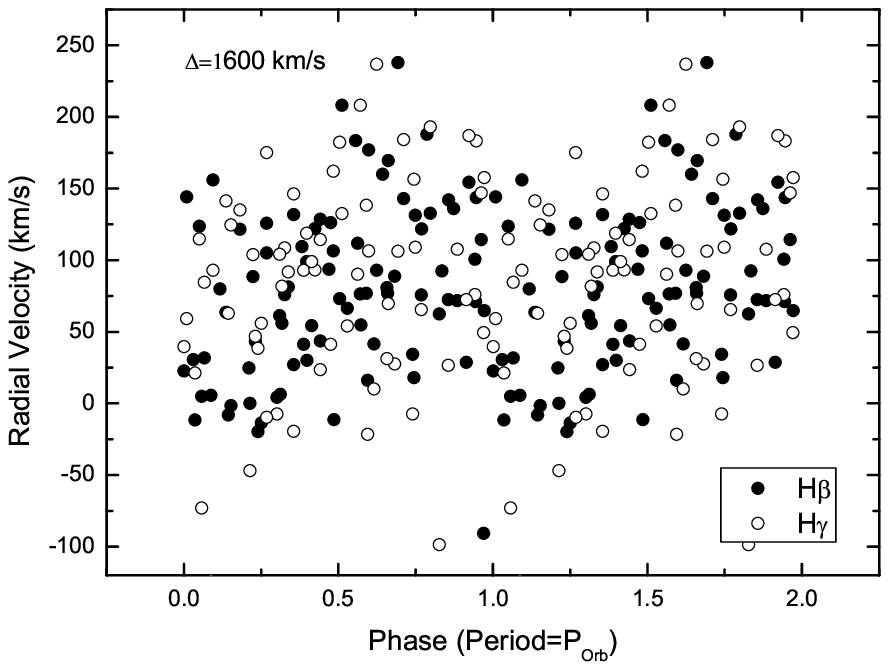}
\plottwo{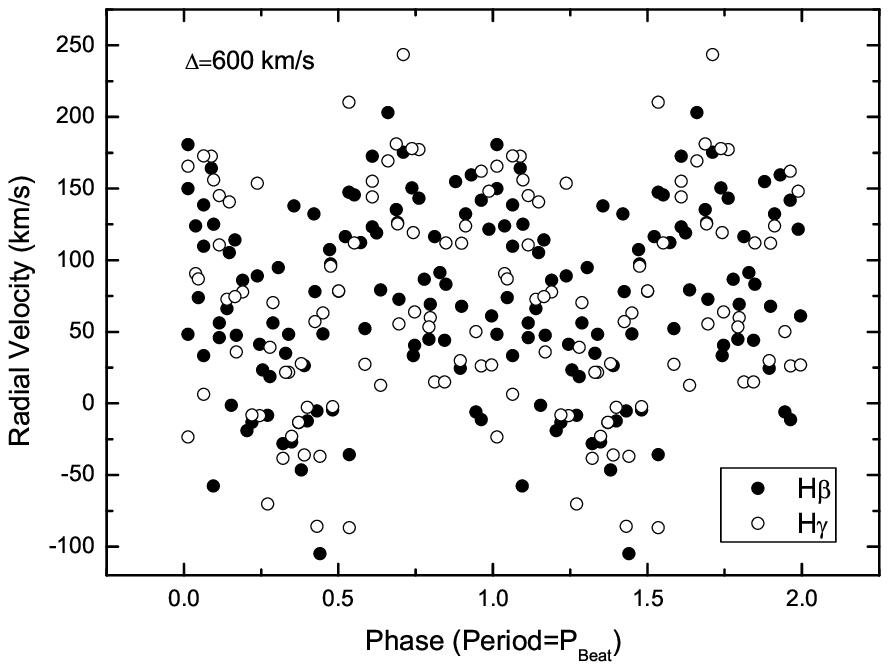}{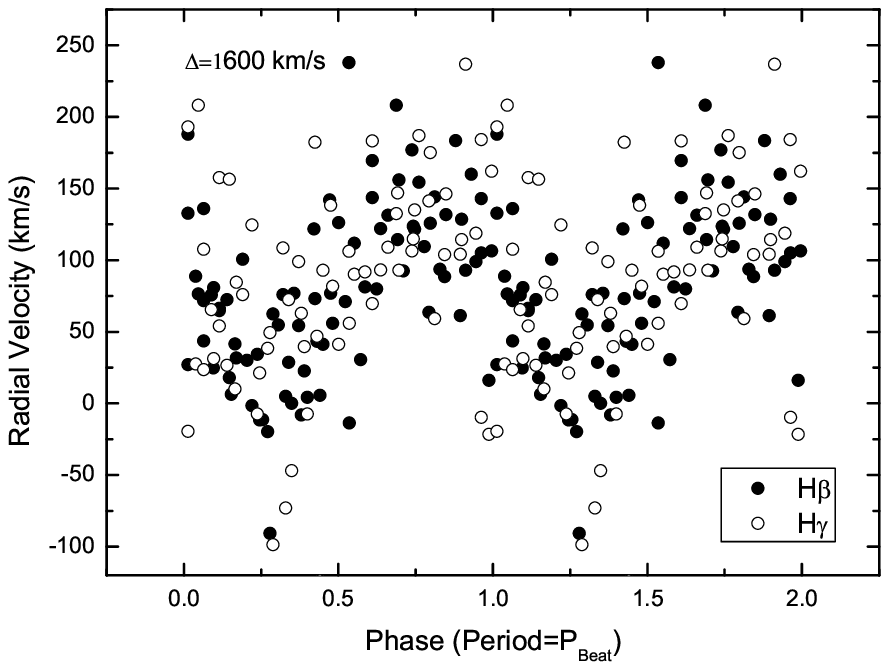}
\caption{The radial velocity curves. On the top left: RV measurements of H$\beta$ and H$\gamma$ lines with double gaussian separation of 600 km/sec folded with the orbital period of FS Aur. On the left bottom are the same measurements folded with the long period. On the left side  corresponding curves
plotted using the  RV measurements with the gaussian separation of 1600 km/sec, corresponding to
800 km/sec velocity in the emission line wings.
\label{rvcurves}}
\end{figure}

Usually in CVs, the line wings at different Gaussian separations show
sinusoidal curves of the same period (orbital), sometimes displaced in phases
relative to each other. The pattern that emerged in case of new observations of FS
Aur was completely different.  Using
the standard Lomb--Scargle power spectrum analysis \citep{Lomb, Scargle},
we calculated the power spectra of the RV measurements with four distinct
separations of the double Gaussians (Fig. 2). At 600 km/sec separation, there
is the power peak at the frequency corresponding to the orbital period. As
the separation increases, the peak corresponding to the orbital period
decreases in strength, while a second strong peak emerges. Frequency of this
peak is equal exactly to the beat between the orbital and 3.425h photometric
periods: 1/Pbeat=1/Porb - 1/Pphot.
By separation of 2000 km/sec, all power is concentrated in the beat period.
The significance levels of the power peaks in Fig \ref{fig2}  far exceed 99.99\% and the result
is repeated with other emission lines (H$\gamma$ and H$\alpha$),
thus leaving no doubt as to its reality.  It must be noted that a peak at the beat
($1/P_{orb} - 1/P_{phot}$) period was detected in our earlier observations
of \object{FS Aur} \citep{Tov1}, which is another indication of
real and repetitive  nature of the discovered phenomenon.
Back then, the object was brighter, which means that the accretion disk was brighter, 
and the modulation in the wings of the line with different period was not so apparent.
The new observations of \object{FS Aur} in a low luminosity state permit detecting 
the produced effect by naked eye. We constructed a trailed spectrum (two
dimensional image of the emission line where the x-axes are
wavelengths or radial velocities and the y-axes are times)  presented in Fig. \ref{fig3}.
The intense central part of the line varies reflecting the orbital motion of the
system, but the  wings of the line show waves with a different period,
i.e. the beat period between the orbital and 3.425h photometric
periods. On the flanks of the trailed spectra we  plotted measurements of
radial velocities of corresponding portions of wings of the line. The RV curve measured 
near the center of the line is presented on the left and measurements in the extreme wings 
are in the utmost right panel.
The radial velocity curves are even more demonstrative when folded with the corresponding periods.
In  Fig \ref{rvcurves}  sets of panels are presented. On the left side the RV measured at 600
km/sec Gaussian separation or within $\pm300$ km/sec from the center are folded with the
orbital period on top and  147\,min beat period on the bottom. On the right side the RV measurements
in far wings of the line at 1600 km/sec separation are folded with the same periods. As
one can see, the center of the line forms the expected sinusoid when folded with the orbital period,
while the wings of the line fold correctly only with the long period. 
When one dataset is folded with the period derived from the second dataset,
then RV curve becomes confused - and vice versa.
In this Figure we also plotted measurements of H$\gamma$ line to
show that pattern repeats from line to line.

The wings of the emission lines in \object{FS Aur} are considerably
fainter than those in \object{HS2331+3905}, in which the double period can be seen
clearly on the trailed spectrogram (see Fig. \ref{fig6}, left panel). This unusual pattern
was first reported by \cite{Araujo05} and persists  in later observations. 
Until now it had been an unique phenomenon reported
without any explanation.  In Fig. \ref{fig6}, we present a sample from
our observations of \object{HS2331+3905}, which vividly demonstrates this
peculiar phenomenon and, as such, is better suited for a comparison
with the model which we developed.

\object{HS2331+3905} is an eclipsing CV, therefore its
inclination is much higher ($i>75^o$) than of \object{FS Aur} for which no exact estimates of
inclination angle exists, but it is believed to be in range of $50^o < i < 65^o$  \citep{Neustroev,Tov1}.
The difference in inclinations probably can explain why the wings of \object{HS2331+3905}
demonstrate a spectacular effect, while for \object{FS Aur} it is barely seen.
However, a more important reason here can be the difference in mass-transfer rates in these systems.
A number of arguments suggest that \object{HS2331+3905} is a CV with a very low mass
transfer rate \citep{Araujo05}, that means the accretion disk in this system is optically
thin and slim. On the contrary, the mass transfer rate in FS Aur is quite high
($\sim4\times10^{10} M_\sun/yr$ - \citealt{Urban_Sion}), thus its accretion disk is expected
to be thick both optically and geometrically. Such  disk can envelope the WD and
obscure central parts, which only can  be seen occasionally through the scattered radiation
in the polar directions of the disk. A decrease of the mass transfer rate could uncover the
innermost disk and allow us to see this area more clearly.
Of course, differences in magnetic field strength and higher spectral resolution during
observations of \object{HS2331+3905} also contribute to the appearance of the emission
lines.

\subsection{Common case and the model}

The evidence for the second period in the wings of \object{FS Aur} emission
lines, directly related to the long photometric period, provided the
crucial link between itself and \object{HS2331+3905}. The parameters of both 
objects, summarised in Tab. \ref{syspar}, are also very similar.
Whilst before, one only could speculate where the long photometric period in 
\object{FS Aur} was originating from, now it is without doubt clear that it is 
associated with the internal parts of the accretion disk.

\begin{figure}
\plotone{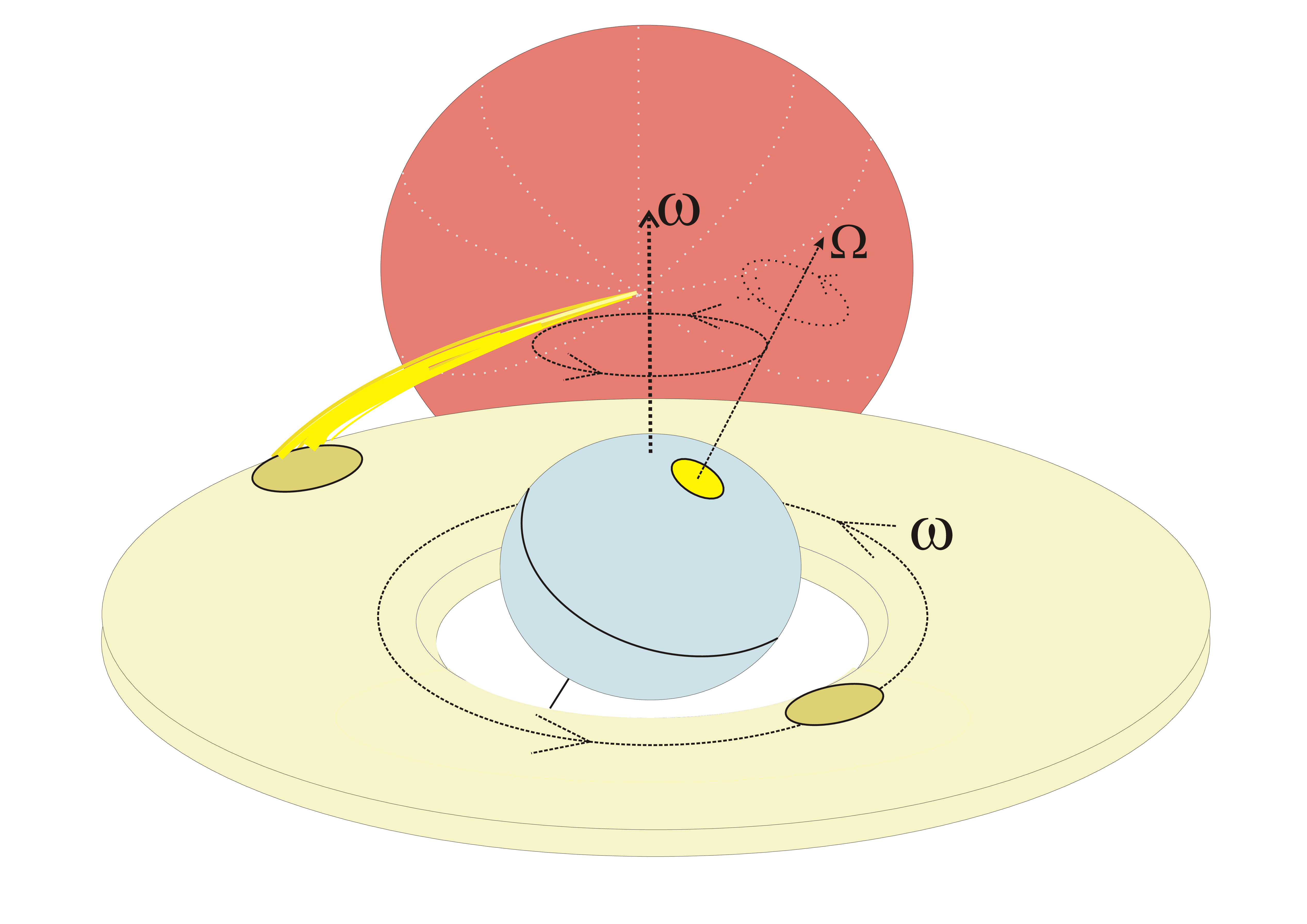}
\caption{A schematic model of a FS Aur type CV with a rapidly
rotating and precessing magnetic white dwarf as a primary, and a truncated accretion disc.
The disc has a heated bright spot inside, which is not stationary, but follows the
rotation axis of the WD as it slowly  precesses. The origin of heating is not clear
at this moment, but can be explained by variety of reasons. There is another bright, hot spot on the outer
edge of the disc, which is stationary and located where the stream of mass transfer
impacts the disc.  \label{kartinka}}
\end{figure}

\begin{figure}
\includegraphics[angle=0,scale=2]{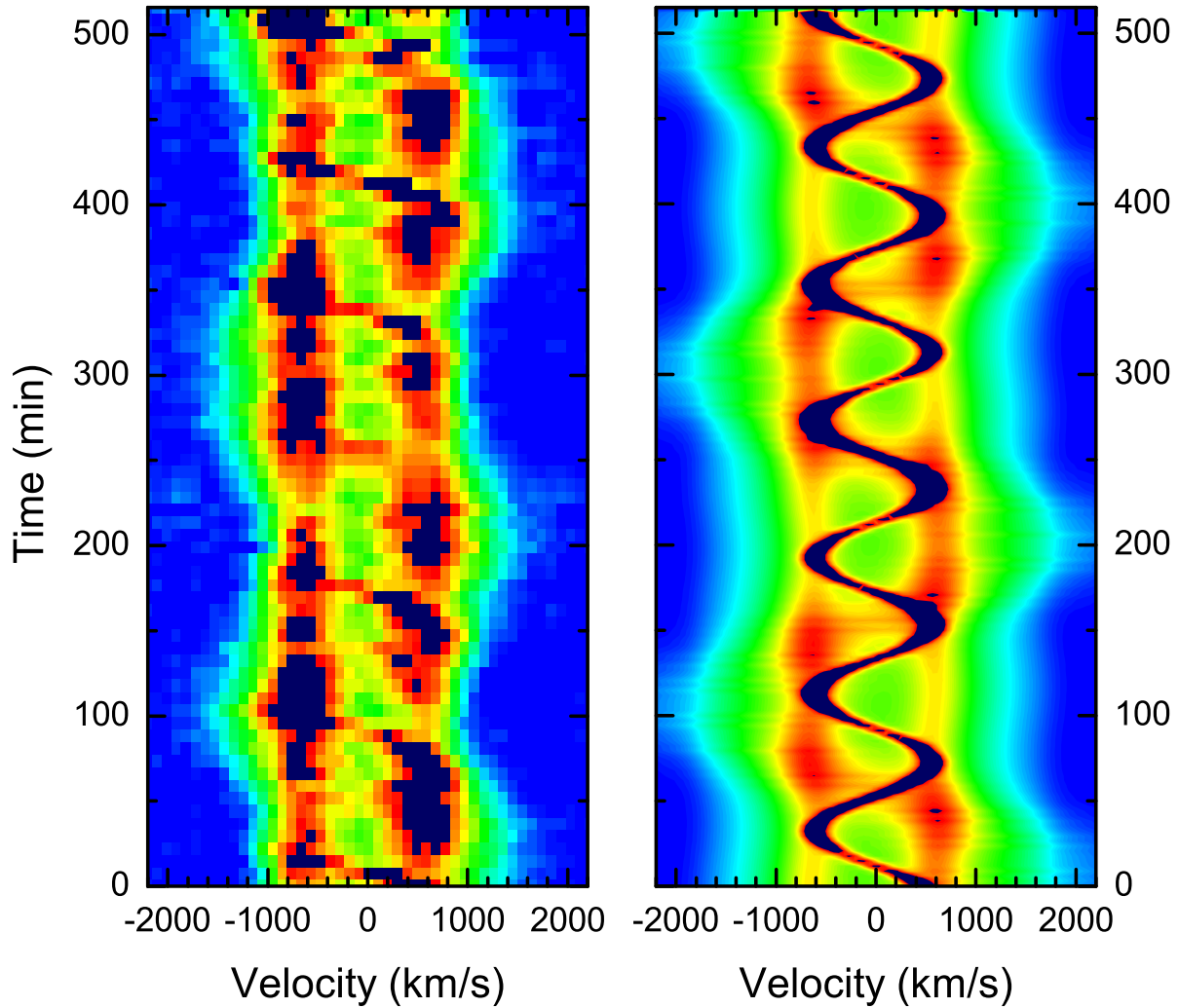}
\caption{Trailed spectrogram of HS2331+3905 H$\beta$ emission line is on the left panel. The
spectra were obtained at the OAN SPM 2.1-m telescope during a multi-longitude
campaign in 2004. The details of the observations and results will appear in
\cite{Gansicke}.
The model describing CV system containing precessing, magnetic and rapidly rotating WD is on the right panel.
We produced this image based upon a three-component model that
includes a geometrically thin Keplerian accretion disk and two bright spots.
\label{fig6}}
\end{figure}

The two objects suggest universal picture, whose common feature, we believe,
is a fast rotating and precessing, magnetically accreting white dwarf.  
We built a simple, basically qualitative model of the accretion disk
that recreates  the observed features of the trailed
spectrogram of these objects, in particular of \object{HS2331+3905}. 
The basic characteristics of the model are:
\begin{enumerate}
\item  The system should be moderately magnetic  (IP) in order to create conditions where the accretion disk exists, but is truncated. More importantly we need an accreting  magnetic pole on the white dwarf that will alter the structure of the inner disk either by irradiation or geometrically (e.g.  warp,  density bulge, etc)
\item Rotation of the spot around the inner radius of the disk in sync with the precessing WD.
\end{enumerate}
 In Fig \ref{kartinka} we present a schematic drawing  of the proposed model.
In order to simulate emission line profiles formed by the
accretion disk of \object{HS2331+3905}, we have applied a three-component model
that includes a geometrically thin Keplerian accretion disk and two
bright spots. The first spot, on the outer rim of the accretion disk,
maintains a constant position with respect to components of the binary
system, while the second spot moves around the inner edge of the disk
to follow the precession of the WD. We began the modeling of the line
profiles with calculation of a symmetrical double-peaked profile
formed in the uniform axisymmetrical disk, then added the distorting
components formed in the bright spots. To calculate the emission line
profiles we divide the disk surface into a grid of elements and
assign the velocity vector, line strength and other parameters for
each element. The computation of the profiles proceeds by summing the
local line profiles weighted by the areas of the surface elements,
taking into account the Keplerian velocity gradient across the finite
thickness of the disk \citep{HM}. We have assumed a
power law function for distribution of the local line emissivity $f(r)$
over the disk surface $f(r )\sim r^{-\alpha}$, where $r$ is the radial distance
from the disk centre and $\alpha = 1- 2.5$ \citep{Smak,Horne1}.
For calculation of the spot's emission components we
consider the spots on the accretion disk to have a quasi-rectangular
shape and a Keplerian velocity. The local line emissivity $f(r)$ over
the spots areas was simply increased by some constant value with
comparison to the underlying accretion disk. For details of the method
see \cite{HM} and \cite{Horne}, whereas its extension for
a non-uniform accretion disk is described by
\cite{Neustroev1}.
On the base of this model we calculated a time dependent
sequence of emission line profiles. The location of the spots, at any
given time, was determined by their current orbital and precession
phase. The first is at the usual place on the outer rim which is common for
most CVs co-rotating with the binary, while the other is on the inner
edge of the disk and whose location is determined by interaction with
the precessing WD. The corresponding trailed spectrogram produced by
this model is presented in Fig \ref{fig6},  right panel.

The inner disk should be warped by the interaction between the magnetosphere 
and the disk, but the heating mechanism is not clear at this point.
Theory predicts a higher density in parts of the disk closest to the magnetic 
pole \citep{Romanova}.
The warped disk as well as the higher density spot will
move and change its angle toward an observer following the precession
axis. There is also the possibility of the irradiation of parts of the
inner disk by the accretion spot on the white dwarf. The irradiation
phenomenon is detected routinely in Polars, where there is no disk,
and the beam from the accretion spot irradiates the facing side of the
secondary star \citep{Warner}.
 As already mentioned the model is
qualitative (or rather geometrical) and does not address the physical
processes responsible for heating or emmisivity of spots or the disk
in general.

\subsection{Why precession of WD and why these two objects?}

With the assumption that there is a  bright spot at the inner side of the accretion disk, 
we were able to mimic the behavior of the emission lines of these two objects. 
But the model is incomplete without a mechanism that makes the inner spot move 
around in a periodic manner, independent of the orbital period. 
In \cite{Tov1} we discussed a variety of  scenarios and suggested that the precession
of  the white dwarf was  the  possible solution, provided that the white dwarf in FS Aur is accreting magnetically and rapidly rotating. The argument was based on a simple premise that such
precession would provide a period comparable to the observed one.
According to \cite{Leins}, the Euler frequency of free precession $\omega_e$ of a rigid
and axially symmetric body or the Chandler frequency $\omega_c$ of an
elastic body will be:
\begin{equation}
\omega_{e,c} = \alpha_{e,c}(\rho_c ) \Omega^3
\end{equation}
where $\alpha_{e,c}$ are coefficients that are determined by the
central density $\rho_c$ and $\Omega$ is the rotation frequency of a
WD.  As a result a WD with spin period in range of 50 to 100 sec should
have a precession period of a few hours\footnote{
Independently, \citet{1992A&A...260..268S}  made rough estimates of $P_{prec}/P_{spin}\sim10^4-10^5$  using slow rotating 
polytropic star models and 
obtained results similar to \cite{Leins}  for long spin periods  typical of majority of IPs, 
but his calculations were not intended for rapid rotators. }, depending on mass of the WD.
Periods observed in \object{HS2331+3905} satisfy spin/precession ratio and strengthen 
our claim that precession of the WD is responsible for observed phenomena.
This also gives more credibility to the ~100 sec period found in the power
spectrum of  \object{FS Aur} in the optical \citep{Neustroev2} and X-rays
\citep{Tov3}.

In \cite{Tov1} we also speculated that the discovered long
photometric period of \object{FS Aur} was the precession period of the WD.
However now, when we found the spectroscopic period at the beat frequency
between orbital  and  photometric, and compare two objects, we should
revise the assignment.  It seems natural that the spectroscopic period will be a direct
measure of rotation of the inner spot (precession period according to our
hypothesis) and that the period seen in photometry will be the beat.
Indeed, there is no way to produce a significant radial velocity signal on
the beat period (at least we were unable to do), but there is no
problem to do this for photometric variations. Following a similar strategy to
that of \cite{1986MNRAS.219..347W}, one can see that any structures which are
co-rotating with the binary with frequency $\omega_{orb}$ and which are
periodically illuminated by the beam as it precesses around the system with
frequency $\omega_{prec}$, will vary in brightness with frequency
($\omega_{orb}$-$\omega_{prec}$)=$\omega_{phot}$.

\begin{figure}
\includegraphics[angle=0,scale=0.7]{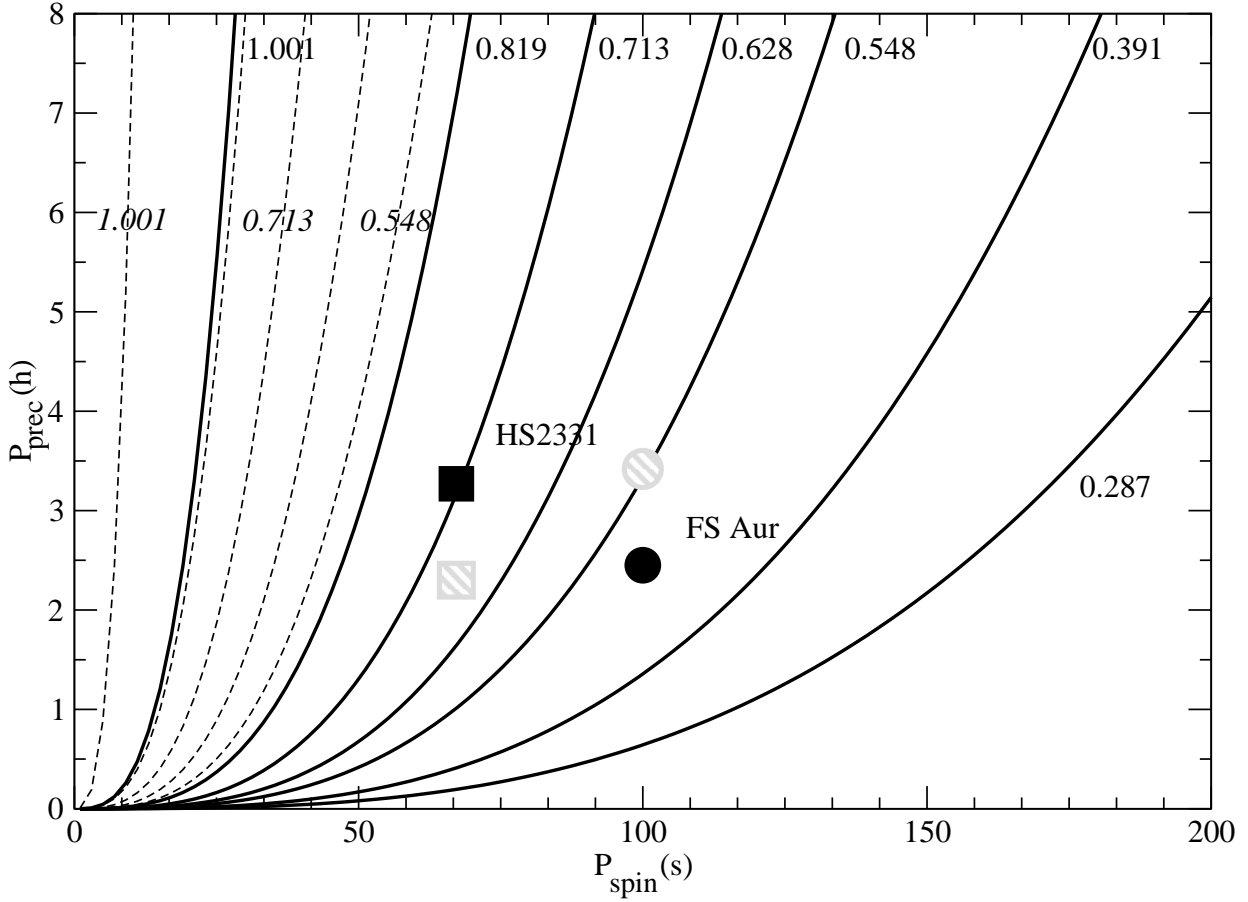}
\caption{The relation $P_{prec}$ vs.  $P_{spin}$  for WDs with different masses
in case of Euler  (solid lines) and Chandler frequency (dashed lines) from Leins et al., (1992).
Numbers mark  the mass of  WD for each curve.
Filled circle and square denote long  spectral periods of FS Aur and HS2331+3905 respectively.
Hashed circle mark long photometric period of FS Aur, which corresponds to the beat period
between orbital and long spectroscopic periods. Respective beat period for HS2331+3905, marked by
hashed square is actually not detected. We assume that spectroscopic, rather than photometric, period corresponds to the precession period,  thus filled symbols should indicate correct positions of the systems in the plot.
\label{fig7}}
\end{figure}

Considering the long spectroscopic period as the precession, we still
reach agreement between the spin and precession periods for  FS Aur
according to  (1). In  Fig \ref{fig7} we present the relationship of the
precession period to the spin period, as follows from the \cite{Leins}
analysis. The solid lines are ratios determined by using the Euler frequency
for a WD with masses indicated by the numbers at the top of the figure.
Corresponding Chandler dependences are shown with dashed lines within the
range of masses indicated lower. The location of FS Aur and HS2331+3905 are
marked by filled symbols, which assumes that the precession period corresponds
to spectroscopic period. If the photometric period is the period of
precession then the symbols shift to the locations marked by the open symbols.
In both cases the masses of the WDs in both systems make perfect sense, given
that they act as rigid bodies and the Euler frequencies are applicable. In
plotting this figure we calculated what the long photometric period of
HS2331+3905 would be as a beat period between observed spin and long spectroscopic
periods, but it has not been detected.

Assuming that the mean white dwarf mass in CVs is of about 0.6 M$_\odot$, only those WDs 
whose spin period is $\leq 120$ sec would have a precession period of less than 8 hours that 
could be detectable in a single night's observing. Detection
of longer periods is not impossible of course, and is a complicated task
and we are not sure anyone ever looked at it. The number of systems that are
proven to be IP with established spin periods is very limitted. One other
known rapid rotator is DQ Her. There are no reports of additional long
periods detected in this system,  but its orbital period is 4.65 hours
(quite different from objects discussed here and compatible with the
hypothetical precession period) and it has a huge accretion disk
\citep{dq}.  Again, we would like to emphasize, that when FS Aur is
observed in normal quiescence, the long period is barely detectable in
the wings of the lines.  So, while the free precession of white dwarfs in
CVs can be a common phenomenon it is not something that is easily detectable.

\section{Conclusions}

We demonstrate that the two enigmatic Cataclysmic Variables \object{FS Aur} and
HS 2331+3905, which were known to show periods much longer than their
orbital period in the photometric light curve and the wings of the
emission lines, respectively, have the same nature. The clarification
derives from the discovery of the second period in the wings of the
accretion lines of \object{FS Aur}, the same phenomenon that was first seen in
\object{HS2331+3905}. This allows us to consider them both as Intermediate
Polars, which, in the case of \object{HS2331+3905}, is an observationally
proven fact. If this is true, then the model of a fast rotating and  precessing,
magnetically accreting white dwarf proposed earlier for  \object{FS Aur},
works for both systems and allows for the explanation of outstanding
features observed in these systems, as demonstrated by the model of an
accretion disk with a bright spot on the inner rim following the
precession axis.  This hypothetical inner bright spot is not well defined, it could be 
the irradiated inner edge of the disk, condensation of matter in it or a warped inner 
disk which are all due to interaction with the magnetosphere. The differences in 
appearance of the long photometric  (beat) period (there was no such period 
reported for \object{HS2331+3905}) and the strength of the modulation in
the wings of the emission lines in the two systems can be
accounted for
by the difference in inclination angles and/or mass-transfer rates. 

If our analysis is correct, then this will be
the first instance when the precession of the WD is detected
in a more or less straightforward manner.
Combined with other
methods to estimate masses and density of the WDs in CVs, the
precession period will help to refine these values, which in turn,
allows for testing and proving the theory for the precession of compact
stars and may allow for further advances in our knowledge of the
structure of WDs and the influence of accretion on them.

\acknowledgments We acknowledge CONACyT grant 45847. VN acknowledges
support of IRCSET under their basic research programme and the support
of the HEA funded CosmoGrid project. We appreciate discussion and
comments made by Dr. M. Romanova regarding the magnetic accretion. We
also acknowledge useful discussion of the subject with
Dr. B. Gansicke. We are thankful to Drs. M. Richer \& J.Tomsick for
careful reading of the manuscript.



\clearpage

\begin{table*}[]
\begin{center}
\caption{Log of observations\label{obslog}}
\begin{tabular}{llllc}
\hline\hline
Object       & Date    & HJD+     & Instrument/Grating$^a$/Range/Exp.Time$\times$N$^b$&   Duration\\
Spectroscopy &         & 2450000  &                                                &            \\ \hline
FS Aur & 2004-12-08 & 3347        & B\&Ch\ \ \ 400l/mm\   3500-5500\AA\  420s$\times$37   & 5.2h  \\
FS Aur & 2004-12-08 & 3347        & B\&Ch\ \ \ 400l/mm\    6470-8500\AA\  420s$\times$17   & 2.4h  \\
FS Aur & 2004-12-09 & 3348        & B\&Ch\ \ \ 400l/mm\     3500-5500\AA\  420s$\times$27   & 3.7h  \\
HS2331+3905 & 2004-09-15 & 3263   & B\&Ch\ \ \ 1200l/mm\   4340-5900\AA\ 420s$\times$67   & 8.6h  \\ \hline
\end{tabular}
\label{tab1}
\end{center}
\begin{tabular}{l}
$^a$ - Boller \& Chivens spectrograph (http://haro.astrospp.unam.mx/Instruments/) \\
$^b$ - Number of Integration \\
\end{tabular}
\end{table*}

\clearpage

\begin{table}
\caption{Parameters of Binary Systems}
\begin{tabular}{lccc}
\hline\hline
     & FS Aur   &   HS\,2331+3905$^2$   \\ \hline
Orbital Period &   85.7 min$^1$     &  81.08 min    \\
Spin Period    &     $\sim 100$ sec$^3$     &   67.2 sec     \\
Long photometric period & 205.5 min$^4$  & -  \\
Long spectroscopic period & 147 min$^5$ & 196.5 min $^5$   \\
Inclination angle & ($50^o-65^o$)$^6$ &  $>75^o$                       \\ \hline
\end{tabular}
\\
\label{syspar}

\begin{tabular}{l}
$1$ - \cite{Thorst},
$2$ - \cite{Araujo04},\\
$3$ - \cite{Neustroev2},
$4$ - \cite{Tov1},\\
$5$ - this paper,
$6$ - \cite{Neustroev}
\end{tabular}

\end{table}

\clearpage



\clearpage

\clearpage


\clearpage

\clearpage

\clearpage

\clearpage

%


\begin{thebibliography}{}
\bibitem[Araujo-Betancor et al. (2004)]{Araujo04}   Araujo-Betancor S.,  GŠNsicke, B. T., Hagen, H.-J.,  Marsh, T.R., Thorstensen, J., Harlaftis, E.T., Fried, R.E.,  Engelset, D., 2004, Revista Mexicana De Astronoma Y Astrofsica  (Serie De Conferencias) 20, 190
\bibitem[Araujo-Betancor et al. (2005)]{Araujo05}  Araujo-Betancor, S.,  G\"{a}nsicke, B.T. ,
           Hagen,  H.-J.,  Marsh, T.R., Harlaftis, E.T.,   Thorstensen, J.,
            Fried, R.F.,   Schmeer, P.,   Engels, D., 2005, A\&A, 430, 629
\bibitem[G\"{a}nsicke et al.(2006)]{Gansicke} G\"{a}nsicke B., et al. 2006, in  preparation
\bibitem[Horne \& Marsh (1986)]{HM} Horne, K.,   Marsh, T.R., 1986, \mnras   218, 761
\bibitem[Horne \& Saar (1991)]{Horne1} Horne, K., Saar, S.H., 1991, \apj, 374, L55
\bibitem[Horne (1995)]{Horne} Horne K., 1995,  A\&A, 297, 273
\bibitem[Lasota \& Hameury (2004)]{Lasota}  Lasota, J.-P.,  Hameury, J.-M. 2004, Magnetic
           Cataclysmic Variables, IAU Colloquium 190, ASP Conference
           Proceedings 315, 46
\bibitem[Leins et al. (1992)]{Leins} Leins, M.,  Soffel, M.H.,   Lay, W.,  Ruder, H.,
           1992, A\&A. 261, 658
\bibitem[Lomb(1976)]{Lomb} Lomb N.~R., 1976, Ap\&SS, 39,
  447
\bibitem[Neustroev(2002)]{Neustroev} Neustroev, V.~V.\ 2002, \aap,
382, 974
\bibitem[Neustroev et al. (2002)]{Neustroev1} Neustroev V.V., Borisov N.V.,
           Barwig H., Bobinger A., Mantel K.H., Simic D., Wolf S.,
           2002, A\&A, 393, 239
\bibitem[Neustroev et al. (2005)]{Neustroev2} Neustroev, V.V.,
           Zharikov, S.V., Tovmassian, G.,  Shearer, A.,  2005,  \mnras ,  362, 1472
\bibitem[Neustroev et al. (2006)]{Neustroev3}Neustroev V.V., et al., 2006 in
           preparation
\bibitem[Norton et al.(1996)]{1996MNRAS.280..937N} Norton, A.~J., 
Beardmore, A.~P., \& Taylor, P.\ 1996, \mnras, 280, 937 
\bibitem[Patterson(2001)]{2001PASP..113..736P} Patterson, J.\ 2001, \pasp, 113, 736
\bibitem[Rodriguez-Gil et al. (2004)]{dwcnc}Rodr'guez-Gil, P.,  G\"{a}nsicke, B. T., Araujo-Betancor, S.,
2004,  MNRAS, 349, 367
\bibitem[Romanova et al. (2004)]{Romanova} Romanova, M.M.,
           Ustyugova, G.V.,  Koldoba, A.V.,   Lovelace, R.V.E., 2004, \apj,  610,
           920
\bibitem[Scargle (1982)]{Scargle} Scargle J.D., 1982, \apj, 263, 835
\bibitem[Schneider \& Young (1980)]{Schneider} Schneider, D.P., Young, P., 1980, \apj,  238, 946
\bibitem[Smak (1981)]{Smak} Smak J., 1981, AcA, 31, 395
\bibitem[Schwarzenberg-Czerny(1992)]{1992A&A...260..268S} Schwarzenberg-Czerny, A.\ 1992, \aap, 260, 268 
\bibitem[Thorstensen et al. (1996)]{Thorst} Thorstensen, J.R., Patterson, J.O., Shambrook, A., Thomas, G. 1996, PASP, 108, 73
\bibitem[Tovmassian et al. (2003)]{Tov1} Tovmassian, G., Zharikov, S., Michel, R., Neustroev, V., Greiner, J.,  Skillman, D. R., Harvey, D. A., Fried, R. E., Patterson, J., 2003, \pasp, 115, 725
\bibitem[Tovmassian (2005)]{Tov2} Tovmassian, G., 2005,  Interacting Binaries: Accretion, Evolution, and
           Outcomes. AIP Conference Proceedings, 797, 257
\bibitem[Tovmassian et al.(2006)]{Tov3} Tovmassian, G. et al., (2006) in
           preparation
\bibitem[Urban \& Sion(2006)]{Urban_Sion} Urban, J.~A., \& Sion, E.~M.\ 2006, \apj, 642, 1029
\bibitem[Warner(1986)]{1986MNRAS.219..347W} Warner, B.\ 1986, \mnras, 219, 
347 
\bibitem[Warner(1995)]{Warner} Warner, B., 1995 in Cataclysmic Variable Stars (Cambridge:Cambridge Univ. Press)
\bibitem[Wood et al.(2005)]{dq} Wood, M.~A., et al.\ 2005, \apj, 634, 570
\end{thebibliography}
\end{document}